# Supervised Segmentation of Retinal Vessel Structures Using ANN

Esra KAYA [1], İsmail SARITAŞ [1], İlker Ali ÖZKAN [2]

[1] *Electrical and Electronics Engineering, Faculty of Technology*
*Selcuk University Alaeddin Keykubat Yerleskesi Selcuklu, Konya, 42075 Turkey*
esrakaya@selcuk.edu.tr
isaritas@selcuk.edu.tr

[2] *Computer Engineering, Faculty of Technology*
*Konya, Turkey*
ilkerozkan@selcuk.edu.tr

*Abstract*—In this study, a supervised retina blood vessel segmentation process was performed on the green channel of the RGB image using artificial neural network (ANN). The green channel is preferred because the retinal vessel structures can be distinguished most clearly from the green channel of the RGB image. The study was performed using 20 images in the DRIVE data set which is one of the most common retina data sets known. The images went through some preprocessing stages like contrast-limited adaptive histogram equalization (CLAHE), color intensity adjustment, morphological operations and median and Gaussian filtering to obtain a good segmentation. Retinal vessel structures were highlighted with top-hat and bot-hat morphological operations and converted to binary image by using global thresholding. Then, the network was trained by the binary version of the images specified as training images in the dataset and the targets are the images segmented manually by a specialist. The average segmentation accuracy for 20 images was found as 0.9492.

## I. Introduction

Due to having visual and perceptive characteristics, eye is one of the most significant parts of human body [1]. Functional disorders occurring in the eye can be diagnosed from vessel like structures in the retina [1]. For example, retinal blood vessels can provide information about pathological diseases such as hypertension and diabetes [2]. Changes in the diameter of the vessel and development of neovascularization are symptoms of diabetic retinopathy, while a decrease in the ratio of the diameter of the retinal arteries to the diameter of the vessels is the indicator of hypertension [3]. Also retinal blood vessels can be examined from outside without requiring intervention because of its visibility [3]. Ophthalmologists diagnose patients by manually examining the status of blood vessels on retinal images which were obtained with high-resolution fundus cameras [2]. The accuracy of manual segmentation is high, but the process can be time-consuming due to the presence of high-volume retinal data [3]. Thus, medical imaging has become an imperative tool in recent years for automatic segmentation instead of manual segmentation [4]. In general, fundus images have low contrast and background noise. Also, the segmentation of blood vessels is difficult because the width, brightness and shape of blood vessels are variant. Thus, it is of great importance to develop an effective image processing technique for an accurate automatic segmentation [3].

Different image processing techniques are used for retinal blood vessel segmentation. These are generally divided into segmentation with supervised and unsupervised learning. Unsupervised methods are rule-based and specify vein regions using the default rules for vessels. Unsupervised learning includes vein tracking, matched filtering, morphological processing, model-based algorithms and multi-scale analysis [3]. Supervised learning is more time-consuming and computationally expensive, as it requires training of complex classifiers that work with large data with different features [3].

Given an example to unsupervised methods, Imani et al. used a morphological component analysis (MCA) algorithm for retinal blood vessel segmentation thus, preventing the formation of false positive vessel pixels in the area of diabetic lesions. As a result, the segmentation accuracy was 0,9524 for the DRIVE data set and 0,9590 for the STARE data set [5]. In another study, the image enhancement realized using mathematical morpholoy and blood vessels were segmented by k-means clustering by Hassan et al. As a result of experimental procedure, 95.10% accuracy was obtained for the DRIVE data set [4]. Frucci et al. carried out the segmentation process using the SEVERE method where the retinal image pixels were assigned to 12 discrete directional information. The accuracy of 0.9550 and 0.9590 were obtained for the test and training images in the DRIVE data set, respectively [6]. Kovács et al. used template matching and contour reconstruction for the self calibration of the retinal blood vessel segmentation system. Accuracy results were found for DRIVE and STARE data sets as 0,9494 and 0,9610, respectively [7]. In another unsupervised segmentation realized by Singh et al., a matched filter based on the function of Gumbel probability distribution was used and the accuracy rate was found 0,9522 for the DRIVE data set and 0,9270 for the STARE data set [2].

As an example of segmentation with supervised method, Soares et al. used 2-dimensional Gabor wavelet for the filtration of the noise of the image thus, improved the vessel appearance.

The accuracy rate obtained using Bayes classifier was 0,9614 [8]. In another study, Ceylan et al. attained the accuracy of 98.56% by using the methods; complex wavelet transform and complex artificial neural network [9]. Vega et al. realized the segmentation process using a lattice neural network with dentritic processing and obtained 0,9412 accuracy for the DRIVE data set and 0,9483 accuracy for the STARE data set [10]. On the other hand, in another supervised segmentation, convolutional neural network (CNN) for feature extraction and random forest (RF) classifier for pixel classification were used by Wang et al. with the aim of realizine a hierarchical segmentation. As a result of the study, the accuracy of 0,9475 and 0,9751 were obtained for DRIVE and STARE data sets, respectively [11]. In another study performed by Ceylan et al., 98.44% accuracy for the DRIVE data set and 99.25% accuracy for the STARE data set were obtained using complex Ripplet-I transformation and complex valued artificial neural network [1]. In this study, a supervised segmentation method using ANN was applied on the green channel of RGB image and the pixels were classified as vessel and non-vessel structures with promising results.

## II. MATERIAL AND METHOD

### A. Image Acquisition

20 original retinal images labeled as training obtained from the DRIVE (Digital Retina Images for Vessel Extraction) data set, which is one of the most commonly used data sets, and 20 binary images corresponding to the training images and segmented by specialists manually were used in this study. Furthermore, DRIVE data set was created from the retinal images acquired from a diabetic retinopathy screening program in The Netherlands by using a Canon CR5 non-mydriatic 3CCD camera with a 45 degree field of view (FOV) [12]. The dimension of the images used in the study was 584 x 565.

### B. Image Preprocessing

Green channel images are usually used as gray level images in retinal segmentation studies due to the visual difference between the blood vesses and the background having the best clarity [13]. In this study, an accurate segmentation of retinal blood vessels has been tried to realize using ANN for a better diagnosis of several diseases such as hypertension and diabetic retinopathy. Fig. 1 shows the original image and the green channel of the first training image in the data set.

Before contrast enhancement for highlighting the vessel like structures, the SNR value, which gives information about the image quality, was measured. SNR value is the signal to noise ratio for each gray image [2]. For images, SNR value can be found by the ratio of the average of image pixel densities to the standard deviation of pixel densities. If the SNR is greater, it means that the detail of the image is more visible. In the study, the contrast parameters have been altered by taking into account the changes in the SNR value to achieve the best contrast value [2]. For contrast enhancement of the images, contrast-limited adaptive histogram equalization (CLAHE) method was used.

Unlike normal histogram equalization, CLAHE works for the optimization of local image contrast by splitting the image into tile like structures. Thus, the contrast can be adjusted for all the regions separately making the details more visible [14]. Also, CLAHE uses a contrast limit known as clip limit for the regions which prevents the noise generation over uniform regions such as background and with this feature, CLAHE differs from normal adaptive histogram equalization [14]. This process is realized by allowing only a maximum number of pixels belonging to the bins related to all local histograms and the pixels which are clipped are distributed uniformly over the whole histogram thus, preserving the former histogram count [14]. In the study, the distribution of CLAHE, meaning the desired histogram shape, was chosen as 'rayleigh' which is a bell shaped histogram and the contrast enhancement limit (clip limit) was chosen as 0.02 [15]. According to the changes made in the study, the recalculated SNR values were found to be higher, thus the details became more apparent. If the clip limit was changed to more or less than this value, it was observed that the SNR value began to deteriorate.

After CLAHE, the density of the image pixels was changed and the color difference between the blood vessels and the background was increased. Apppplication of median filtering on contrast enhanced images made it possible to remove salt-pepper noise and vessel like structures were made more distinct by applying 2-dimensional Gaussian distributed matched filter of which the function is shown below [13]. The images obtained after CLAHE and filtering process are shown in Fig. 2 respectively.

$$G(x,y) = \frac{1}{2\pi\sigma^2} exp(-\frac{x^2+y^2}{2\sigma^2}) \qquad (1)$$

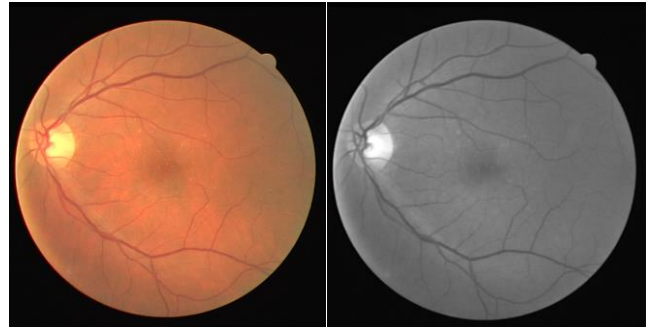

**Fig. 1.** First DRIVE training image and its green channel

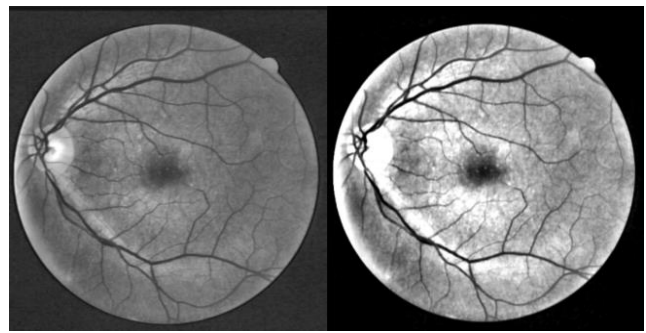

**Fig. 2.** The images after CLAHE and filtering processes

Furthermore, top-hat morphological processing has been used to make blood vessels more prominent [13]. With the use of bottom-line morphological processing and subtraction of these two images from each other also resulted in blood vessels being separated from the background. After these processes, the images are converted to binary system images and the complement of the images were taken. After complementing the images obtained by the specialists, the images were ready for training with using ANN. The image after morphologic operations, the complemented binary image and ground truth which is the image manually segmented by the specialists are shown in Fig. 3, respectively.

### C. Artificial Neural Network

Artificial Neural Networks (ANN) solves the problems in the real world as an impersonation of human brain [16]. An ANN consists of three sequential and connected layers which are input layer, hidden layer and output layer [16, 17]. The data admitted by the input layer is processed by the hidden layer relying on the connections to the input layer and the weights belonging to these connections and then it is transmitted to the output layer which categorizes the data into different groups [16]. The ideal number of neurons in the hiiden layer is chosen by trial and error because there is no analytical method for its determination [16, 17].

In the study, the pixel densities of the binary form of the first training image was chosen as input data for the ANN and the pixel densities of the binary form of the manually segmented image orresponding to the first training image was chosen as target data for the ANN. The data was chosen randomly as training, test and validation data by the percentage of 90, 5 and 5, respectively. The ANN was chosen as a pattern recognition network which is feedforward network with 10 hidden neurons, resilient backpropagation training function and hyperbolic tangent sigmoid transfer function. After the training process, the trained network which has an overall 96.8% accuracy through the training, test and validation data was converted to a function to be operated on the all images in the database. The confusion matrix of the ANN is shown in Fig. 4, the performance of the ANN is shown in Fig. 5 and the receiver operator characteristics of the ANN is shown in Fig. 6.

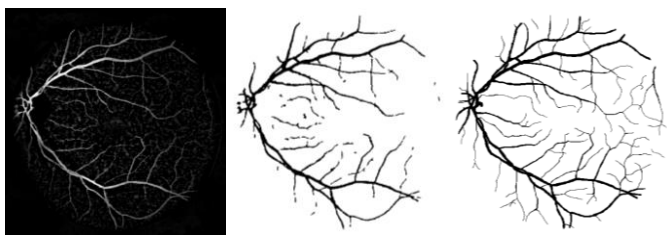

**Fig. 3.** The image after morphological operations, complemneted binary image and ground truth

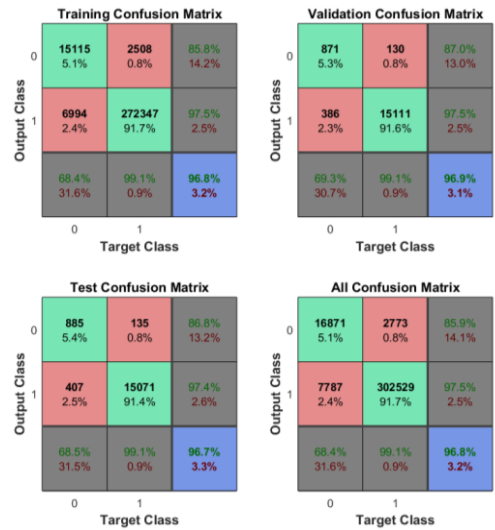

**Fig. 4.** Training, Test, Validation and All Confusion Matrix

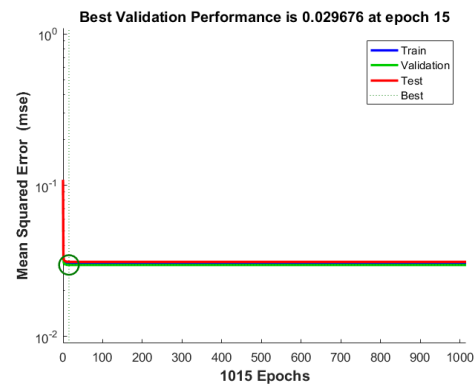

**Fig. 5.** The performance graph of the ANN

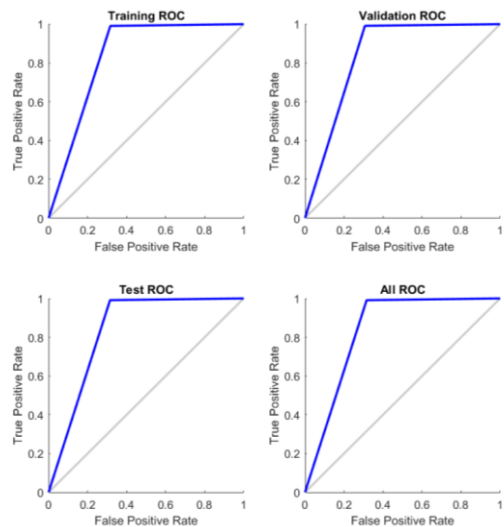

**Fig. 6.** Training, Test, Validation and All ROC

III. RESULTS AND DISCUSSIONS

The function generated from the ANN was applied to the complemented binary images in the database and the results were compared with images segmented by the specialists manually. For the calculation of general accuracy of all images, concepts such as true positive (TP), true negative (TN), false positive (FP) and false negative (FN) were used to evaluate the pixels. True positive refers to the classification of vessel pixels as vessels while true negative refers to the classification of background pixels as background. Also, false positive refers to the classification of background pixels as vessel and false negative refers to the classification of vessel pixels as background.

The accuracy rate is given by the following formula.

$$Accuracy = (TP + TN)/(TP + TN + FP + FN) \quad (2)$$

According to the results obtained based on these values, the average accuracy rate of the segmentation for the training images in the database was found to be 0.9492. In addition, the segmentation process took very short time which is a total of 24.4826 seconds for the total of 20 images, although the supervised segmentation takes a lot of time because of its complexity. The results showed that segmentation was very succesfull and close to an expert's work and took very little time for the total of 20 images which can be minutes for only one image if it was segmented manually. In Table 1, the results of our study are given for comparison with a mixture of other studies involving unsupervised and supervised segmentation methods in the literature. Table 2 shows the accuracy results for all the images in the study.

**Table 1**
Literature Studies of Supervised and Unsupervised Segmentation

| Author | Year | Method | Accuracy |
|---|---|---|---|
| Ceylan et al. | 2013 | Complex Wavelet & Complex Valued ANN | 0,9856 |
| Ceylan et al. | 2016 | Complex Ripplet-I Transform & Complex Valued ANN | 0,9844 |
| Soares et al. | 2006 | 2-D Gabor Wavelet & Bayes Classifier | 0,9614 |
| Frucci et al. | 2016 | SEVERE | 0,9570 |
| İmani et al. | 2015 | Morphological Component Analysis | 0,9524 |
| Singh et al. | 2016 | Gumbel Probability Distributed Matched Filter | 0,9522 |
| Hassan et al. | 2015 | Mathematical Morph. & k-means clustering | 0,9510 |
| Kovács et al. | 2016 | Template Matching & Contour Reconstruction | 0,9494 |
| **This study** | **2017** | **CLAHE & Gaussian Matched Filter & ANN** | **0,9492** |
| Wang et al. | 2015 | Convolutional Neural Network & Random Forest Classifier | 0,9475 |
| Vega et al. | 2015 | Lattice Neural Network with Dentritic Processing | 0,9412 |

**Table 2**
Accuracy Values Obtained for All Images (Average Accuracy is 0.9492)

| DRIVE Images | Accuracy | DRIVE Images | Accuracy | DRIVE Images | Accuracy | DRIVE Images | Accuracy |
|---|---|---|---|---|---|---|---|
| 21_training.tif | 0,9680 | 26_training.tif | 0,9291 | 31_training.tif | 0,9579 | 36_training.tif | 0,9447 |
| 22_training.tif | 0,9574 | 27_training.tif | 0,9575 | 32_training.tif | 0,9627 | 37_training.tif | 0,9505 |
| 23_training.tif | 0,9116 | 28_training.tif | 0,9551 | 33_training.tif | 0,9629 | 38_training.tif | 0,9571 |
| 24_training.tif | 0,9476 | 29_training.tif | 0,9428 | 34_training.tif | 0,9158 | 39_training.tif | 0,9544 |
| 25_training.tif | 0,9523 | 30_training.tif | 0,9457 | 35_training.tif | 0,9555 | 40_training.tif | 0,9545 |

IV. CONCLUSION

Evaluation of the results showed that the segmentation process was successful, not exhausting and took a little time. It can be seen that the procedure will make the specialist's job easier and they can serve more patients without dealing with manual segmentation. In the future, different contrast enhancement, filtering, or morphological operations will be attempted to enhance the image and, different features and supervised learning techniques will be used for comparison and improving the segmentation results.

REFERENCES

[1] Ceylan, M. and H. YAŞAR, *A novel approach for automatic blood vessel extraction in retinal images: complex ripplet-I transform and complex valued artificial neural network.* Turkish Journal of Electrical Engineering & Computer Sciences, 2016. **24**(4): p. 3212-3227.
[2] Singh, N.P. and R. Srivastava, *Retinal blood vessels segmentation by using Gumbel probability distribution function based matched filter.* Computer methods and programs in biomedicine, 2016. **129**: p. 40-50.
[3] Aslani, S. and H. Sarnel, *A new supervised retinal vessel segmentation method based on robust hybrid features.* Biomedical Signal Processing and Control, 2016. **30**: p. 1-12.
[4] Hassan, G., et al., *Retinal blood vessel segmentation approach based on mathematical morphology.* Procedia Computer Science, 2015. **65**: p. 612-622.
[5] Imani, E., M. Javidi, and H.-R. Pourreza, *Improvement of retinal blood vessel detection using morphological component analysis.* Computer methods and programs in biomedicine, 2015. **118**(3): p. 263-279.
[6] Frucci, M., et al., *Severe: Segmenting vessels in retina images.* Pattern Recognition Letters, 2016. **82**: p. 162-169.
[7] Kovács, G. and A. Hajdu, *A self-calibrating approach for the segmentation of retinal vessels by template matching and contour reconstruction.* Medical image analysis, 2016. **29**: p. 24-46.
[8] Soares, J.V., et al., *Retinal vessel segmentation using the 2-D Gabor wavelet and supervised classification.* IEEE Transactions on medical Imaging, 2006. **25**(9): p. 1214-1222.
[9] Ceylan, M. and H. Yacar. *Blood vessel extraction from retinal images using complex wavelet transform and complex-valued artificial neural network.* in *Telecommunications and Signal Processing (TSP), 2013 36th International Conference on.* 2013. IEEE.
[10] Vega, R., et al., *Retinal vessel extraction using lattice neural networks with dendritic processing.* Computers in biology and medicine, 2015. **58**: p. 20-30.
[11] Wang, S., et al., *Hierarchical retinal blood vessel segmentation based on feature and ensemble learning.* Neurocomputing, 2015. **149**: p. 708-717.
[12] Staal, J., et al., *Ridge-based vessel segmentation in color images of the retina.* IEEE Transactions on Medical Imaging, 2004. **23**(4): p. 501-509.
[13] Zhu, C., et al., *Retinal vessel segmentation in colour fundus images using Extreme Learning Machine.* Computerized Medical Imaging and Graphics, 2017. **55**: p. 68-77.
[14] Zuiderveld, K., *Contrast limited adaptive histogram equalization*, in *Graphics gems IV*, S.H. Paul, Editor. 1994, Academic Press Professional, Inc. p. 474-485.
[15] Mathworks. *Contrast-Limited Adaptive Histogram Equalization*. 2017 [cited 2017 10.08.2017]; Available from: https://www.mathworks.com/help/images/ref/adapthisteq.html.
[16] Ebrahimi, E., K. Mollazade, and S. Babaei, *Toward an automatic wheat purity measuring device: A machine vision-based neural networks-assisted imperialist competitive algorithm approach.* Measurement, 2014. **55**: p. 196-205.
[17] Jagtap, J. and M. Kokare, *Human age classification using facial skin aging features and artificial neural network.* Cognitive Systems Research, 2016. **40**: p. 116-128.